\documentclass[prl,twocolumn,preprintnumbers,amsmath,amssymb]{revtex4-1}

\usepackage{graphicx}% Include figure files
\usepackage{dcolumn}% Align table columns on decimal point
\usepackage{bm}% bold math
\usepackage{mathrsfs}
\usepackage{appendix}
\usepackage{bbm}
\usepackage{color}

\newcommand{\be}{\begin{equation}}
\newcommand{\ee}{\end{equation}}
\newcommand{\bea}{\begin{eqnarray}}
\newcommand{\eea}{\end{eqnarray}}

\begin{document}
%\draft
%\preprint{\today }
%
% title page
%

\title{Photon Counting as a Probe of Superfluidity in a Two-Band Bose Hubbard System Coupled to a Cavity Field}

\author{Sara Rajaram and Nandini Trivedi\\
\textit{\small{The Ohio State University, 191 W. Woodruff Avenue, Columbus, OH 43210, USA.}}} 

\date{\today}

\begin{abstract}

We show that photon number measurement can be used to detect superfluidity for a two-band Bose-Hubbard model
coupled to a cavity field. The atom-photon coupling induces transitions between the two internal atomic levels and results in entangled polaritonic states. 
In the presence of a cavity field, we find different photon numbers in the Mott-insulating versus superfluid phases, providing a method of distinguishing the atomic phases by photon counting. Furthermore, we examine the dynamics of the photon field after a rapid
quench to zero atomic hopping by increasing the well depth.
We find a robust correlation between the field's quench dynamics and the initial superfluid order parameter, thereby providing a novel and accurate method of determining the order parameter.  

\end{abstract}

\maketitle

%1)DISCUSS THOMPSON WORK
%2)GO THROUGH MATH in QUENCH DYNAMICS TO MAKE SURE THERE ARE NO OTHER AMP/Freq contributions
%3) READ RECENT RITSCH REVIEW PAPER
%4) INCLUDE vs g equilibrium work (both MF and ED). Check how high of g we can go to in reality
%5) REDO MF EQ PLOTS TO TAKE OUT the ED PART

\textit{Introduction.--} Optical lattices provide very clean and highly tunable testing grounds of important strongly correlated Hamiltonians
and quantum phase transitions.
In particular, numerous experimental groups realized the bosonic Mott-insulating (MI) to superfluid (SF) phase transition in an optical lattice \cite{GreinerQPT}, predicted by the Bose-Hubbard Hamiltonian \cite{Jaksch}. However, the measurement process still stands to be an experimental challenge because current methods to probe these phases rely on destructive time-of-flight measurements.  To detect superfluidity, experimentalists look for peaks in the resulting interference pattern. However, numerical simulations \cite{Nandini} have shown that peaks are not conclusive proof of superfluidty because they may exist even above the critical temperature. 
%In addition, an interference pattern does not offer a quantitatively accurate method of deducing the order parameter, a very important quantity for characterizing the transition.

Fortunately, experimentalists can acquire time-resolved photon statistics with high precision. Therefore one possibility is to circumvent these roadblocks by coupling the atomic system to light. One could then probe the atomic phases by the imprint they leave upon the emitted light. Recent advances in engineering strongly coupled cavity QED systems provide the opportunity to explore fundamental light-matter interactions at the quantum level \cite{Haroche, Colombe, Esslinger, Thompson} . We are encouraged by such experimental progress to consider harnessing the light-matter entanglement to probe novel quantum phases. Here we provide a robust nondestructive method to both distinguish the SF and MI phases and determine the order parameter.

To meet our ends, we extend the Bose-Hubbard model to include two species of bosons with nearest neighbor tunneling. These two species are assumed to correspond to internal levels of a single boson, and interaction with the photon field induces transitions between the two levels \cite{JC}. This model can be viewed as an extension of the Dicke model, within the rotating wave approximation, to bosons in a lattice system \cite{Dicke}.

As in previous works \cite{Bhaseen09, Silver10, Ritsch}, we find a rich phase diagram for this system, including two component superfluidity and Mott-insulating states.  However, previous works \cite{Bhaseen09, Silver10} approximate the cavity field to be coherent, whereas we do not make this assumption. As a result, we find differing shapes and positions of the phase boundaries when light is present. However, if the field is truly a coherent state, the intracavity photon number would remain constant after a quench because the field and atoms are not entangled. Exact diagonalization produces oscillatory behavior, thereby suggesting the coherent field treatment misses a piece of the picture. We explore these oscillations by a theoretical analysis that allows light-matter entanglement. As we will show, these oscillations are a useful probe of the order parameters. 
%{\red{SARA needs some clarification}}

 Our main results are as follows. 1) The average photon number is lower in the SF phase than in the MI phase. Thus, photon counting is a conclusive method to differentiate the atomic phase. 2) The entanglement leads to nontrivial photon number oscillation after a quench. The amplitudes of the oscillations increase with the initial value of the atomic tunneling parameter and correspondingly the initial order parameter. Thus, it is possible to deduce the initial order parameter by recording the photon number dynamics after the quench. 

\textit{Model.--} Let $H_a$ and $H_b$ denote the bare Hamiltonians of the atomic species. Let $H_{field}$ denote the energy contributions from the photon field and the photon-atom coupling. Then, the complete system Hamiltonian, assuming negligible cavity dissipation and spontaneous emission, is given by: 

\begin{figure} [t!]
\centerline{\includegraphics[width=0.35 \textheight]{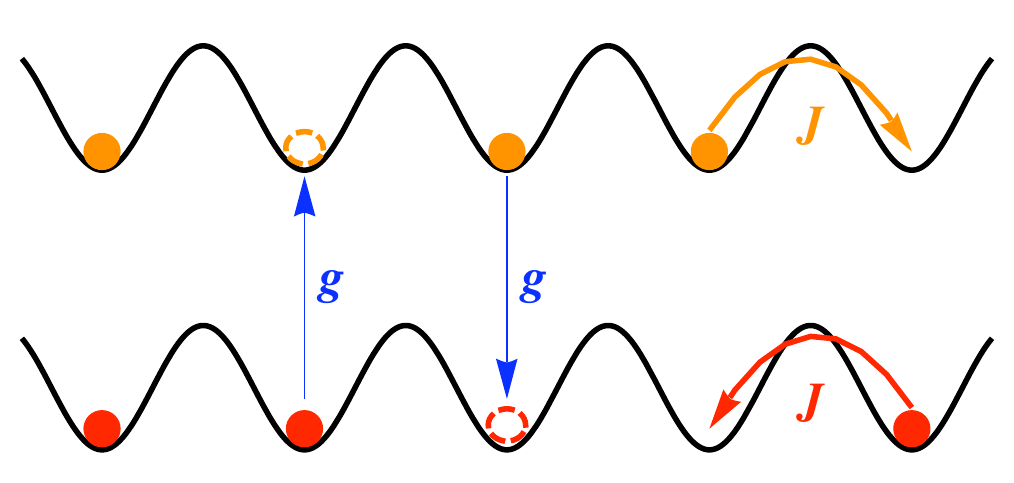}}
\caption{
		Schematic of Hamiltonian, where the cavity field and walls are omitted for clarity. The system consists of bosonic atoms with two internal levels. The atoms in their upper level are depicted in orange while those in the lower level are in red. The cavity field allows transitions between the two internal levels, as shown by blue arrows; $g$ denotes the strength of the atom-photon coupling. The arrows with parameter strength $J$ indicate atomic tunneling to nearest neighbor sites. For simplicity, we consider only hardcore intra-species interactions and zero inter-species interaction.  
	\label{schematic}}
\end{figure}

\be
H = H_{a} + H_{b} + H_{field}- \mu_1 N_1 - \mu_2 N_2
\ee
where,
\begin{align}
H_{a}&= -J\sum_{\left<ij\right>}(a^\dagger_i a_j + h.c.) +  \epsilon_{a}\sum_{i}\hat{n}_{i}^{a}\\ 
H_{b}&= -J\sum_{\left<ij\right>}(b^\dagger_i b_j + h.c.) +  \epsilon_{b}\sum_{i}\hat{n}_{i}^b \\
H_{field}&=\omega \psi^{\dagger} \psi + \frac{g}{\sqrt{N}} \sum_{i} (b_{i}^{\dagger} a_{i} \psi +  b_i a_i^{\dagger} \psi^{\dagger})
\end{align}

Here, $a_i$ and $b_i$ are the respective ground state and excited state bosonic annihilation operators at site $i$, and $\psi$ is the annihilation operator of the cavity field. The on-site field-atom coupling, $\frac{g}{\sqrt{N}} (b_{i}^{\dagger} a_{i} \psi +  b_i a_i^{\dagger} \psi^{\dagger})$, allows for atomic transition to the excited state by absorption of a photon and the reverse process of emission. The parameter $J$ is the atomic nearest neighbor tunneling amplitude, assumed to be the same for both species. For simplicity, we consider hardcore intra-species interactions, and zero inter-species interaction. However, our main results do not qualitatively depend sensitively upon the strength of inter-species interaction.

Finally, we work in the grand canonical ensemble with two chemical potentials corresponding to two conserved quantities: the total number of atoms and the total number of excitations. The total number of atoms is $N_1= \sum_i n_i^a+n_i^b$, 
controlled by the chemical potential, $\mu_1$. Following reference \cite{Silver10}, the total number of excitations is $N_2=\sum_i \psi_i^{\dagger} \psi_i +\frac{1}{2}(n_i^b-n_i^a +1)$, controlled by its chemical potential $\mu_2$.

We are primarily interested in the dependence of the system on $J/g$, as we want to focus on the relation between the atomic phases and the light-matter coupling. %Thus, we fix many of the other parameters throughout this paper. Foremost, we prefer $g$ to be above the super-radiant transition, yet small enough that the two branches of the n-photon polaritons are close in energy. As we will see, this choice leads to interesting photon dynamics. Thus, we choose $g=1$.
Thus, we choose $g$ to be our unit. To keep consistency and allow comparisons with previous works \cite{Bhaseen09, Silver10}, we also fix the following parameters: $\mu_1/g=-.6$ and $\delta/g=1$, where $\delta=\epsilon_b-\epsilon_a-\omega$. 

\begin{figure} [t!]
\centerline{\includegraphics[width=3.2 in]{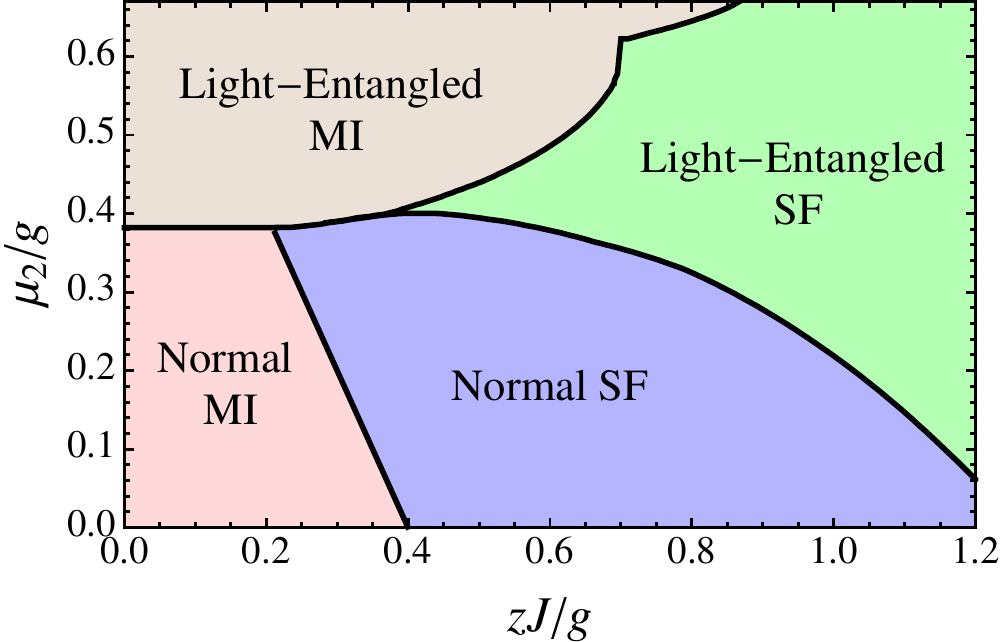}}
\caption{The phase diagram shows four distinct phases in the $\mu_2/g-J/g$ plane. Here $\mu_2$ controls the number of excitations, $J$ is the inter-site tunneling rate, and $g$ is the atom-photon coupling.
The lower half of the phase diagram has no excitations, leading to a typical MI-SF transition of uni-species hardcore bosons. For larger $\mu_2$, we predict a field-matter entangled system. As one tunes $J/g$ in this region, there is a MI-SF transition of both atomic species.}
\label{phasediag}
\end{figure}

\begin{figure*} [t!]
\centerline{\includegraphics[width=7.1 in]{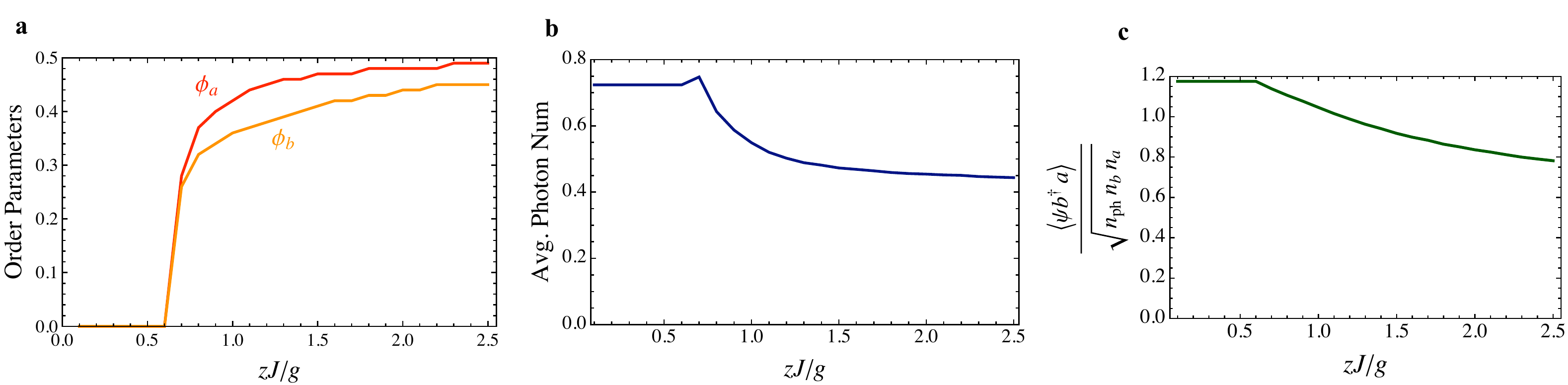}}
\caption{
		a) Plot shows how the mean field order parameters of the ``a atoms"  and ``b atoms" vary with $zJ/g$, indicating a first order transition when the system is light entangled. b)  Photon number density, $\langle n_{ph} \rangle$, as a function of hopping strength, $zJ/g$. c) Measure of photon absorption, $| \langle  \psi_i b_i^{\dagger} a_i \rangle|/\sqrt{n_{ph} n_b n_a}$, as a function of hopping strength, $zJ/g$. Plots a-c indicate that the Mott insulating to superfluid phase transition is accompanied by a change in both the intracavity photon number and the photon absorption strength. 
	\label{eqplots}}
\end{figure*}

To make the atomic tunneling terms more tractable, we employ a mean field approximation. Following known methods \cite{Sheshadri, Demler}, we arrive at the mean field Hamiltonian ($H_{MF}$): 

\bea \label{meanfieldham}
H_{MF}&=&\sum\limits_{i}   H_{MF}^a+H_{MF}^b+\omega \psi_{i}^{\dagger} \psi_{i} + \\ \nonumber
&&  + g (b_{i}^{\dagger} a_{i} \psi_{i} + b_i a_i^{\dagger} \psi_{i}^{\dagger}) - \mu_{1} n_{i1} - \mu_{2} n_{i2} 
\eea
with 
\begin{align}
H_{MF}^a &=  \epsilon_{a} n_{ia} - zJ \phi_{a}(a_{i}^{\dagger} + a_{i})+ zJ|\phi_{a}|^{2} \\
H_{MF}^b &=\epsilon_{b} n_{ib} - zJ \phi_{b}(b_{i}^{\dagger} + b_{i})+zJ|\phi_{b}|^{2} 
\end{align}
where we define $\phi_a=\langle a \rangle$ and $\phi_b=\langle b \rangle$, the respective superfluid order parameters for the lower and upper level atoms. 

Note that we also consider $\psi_i$ to be a local operator, rather than global. To justify this, note that $\psi_k=\frac{1}{\sqrt{N}}\sum_{\mathbf{r_i}}e^{i\mathbf{k} \cdot \mathbf{r_i}} \psi_i$.  However, $|\mathbf{k} \cdot \mathbf{r_i}| $ is negligibly small, allowing the replacement $\psi_k \rightarrow \sum_{\mathbf{r_i}} \psi_i$. The following sections exhibit the results of our mean field assumptions. However, we note that the trends are in qualitative agreement with exact diagonalization results. 

\textit{Equilibrium Phases.--} We numerically minimize $H_{MF}$ with respect to $\phi_a$ and $\phi_b$ to determine the equilibrium phases in the $\mu_2/g-J/g$ plane. See Figure \ref{phasediag} for the phase diagram. In the light-entangled regime, $H_{MF}$ predicts a first order phase transition (Figure \ref{eqplots}a) between a MI and a two component SF state. Within mean field, the MI state is a lower branch polariton state on each site. This is clear from setting $\phi_a=\phi_b=0$ in $H_{MF}$ and then noting that $H_{MF} \rightarrow \sum_i H_i^{JC}$, where $H_{i}^{JC}$ is simply the Jaynes-Cummings Hamiltonian on site $i$. However, the superfluid phase is a delocalized state of matter, thereby necessarily a linear combination of multiple polariton states. 

Looking at Figure \ref{eqplots}b, we also see that the MI to SF phase transition is associated with a discontinuous drop in the photon number density, and the number decreases continuously as we raise $J/g$ in the SF regime. This behavior results from nonlocal hopping taking precedence over the local light-atom interaction with increasing $J/g$. Thus, the photon number tracks the atomic phase, so photon number measurement is an easy experimental method to accurately detect superfluidity.

\textit{Quench Results--} The above features as well as past works \cite{RitschPRA, RitschNature} provide methods to distinguish the two atomic phases. However, these works do not distinguish the degree of superfluidity. To do so, we need a quantity that varies sensitively with $J$. Looking at  Figure \ref{eqplots}c, we see that $| \langle \psi_i b_i^{\dagger} a_i \rangle|$, a measure of the strength of photon absorption, does in fact change sensitively with $J$. Unfortunately, $| \langle \psi_i b_i^{\dagger} a_i \rangle |$ is not a directly measurable quantity. We therefore consider the equation of motion of the intracavity photon number. The number dynamics directly depends on $\psi_i b_i^{\dagger} a_i$ according to,

\begin{figure*} [t!]
\centerline{\includegraphics[width=1.1 \textwidth]{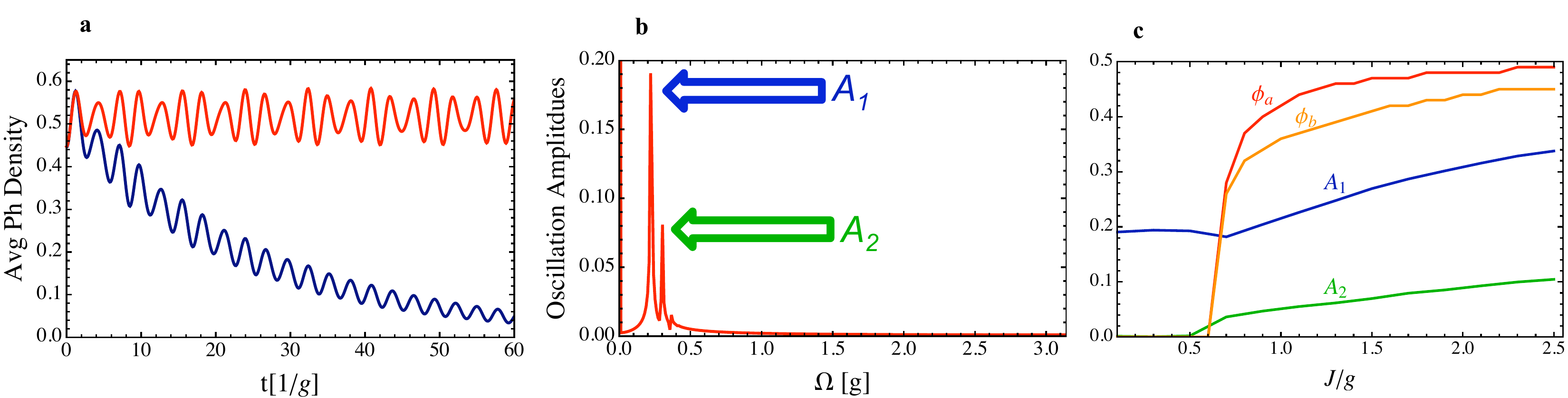}}
\caption{
		a) Quench dynamics of the intracavity photon number density, found from time evolving an initial superfluid state. The red curve assumes zero dissipation. The blue curve accounts for photon leakage by averaging over 10000 wave function Monte Carlo realizations ($\frac{\kappa}{g}=.05$). The dissipation modifies the photon dynamics by causing an overall decay, but the oscillation is preserved for sufficiently small $\kappa$. b) Fourier transform of the zero-dissipation-dynamics indicates two dominant frequencies. The corresponding amplitudes of photon oscillation are denoted $A_1$ and $A_2$. c) Figure shows the amplitudes, $A_1$ and $A_2$, in the presence of dissipation for varying initial hopping. We see that the amplitudes closely correspond to the superfluid order parameters. This trend offers a novel experimental method to deduce the order parameters from photon number dynamics alone. Note: for the plots obtained from Monte Carlo, error bars are smaller than the plot line width.  
	\label{timeEvolPh}}
\end{figure*}

\be
\frac{d(\psi_i^{\dagger} \psi_i)}{dt}= ig(\psi_i b_i^{\dagger} a_i - \psi_i^{\dagger} b_i a_i^{\dagger})
\ee

Furthermore, the time evolution of $ \psi_i b_i^{\dagger} a_i $ depends on the initial order parameters, $\phi_a$ and $\phi_b$:
\begin{align}
\frac{d(\psi_i b_i^{\dagger} a_i )}{dt}&=i[H_a, \psi_i b_i^{\dagger} a_i ] + i[H_b, \psi_i b_i^{\dagger} a_i ] + ...\\
\frac{d(\psi_i b_i^{\dagger} a_i )}{dt}&= i zJ\phi_a \psi_i b_i^{\dagger} +...+- i zJ\phi_b \psi_i a_i + ... 
\end{align}

 Therefore the absorption strength is the bridge between an observable, the photon number dynamics, and the desired quantity, the superfluid order parameters. 
 Here we propose a way of accessing the information contained in the photon field's time evolution. 

 Our method exploits the quench dynamics to capture how the field-matter entanglement affects the photon number dynamics. Consider a system in the light-entangled superfluid regime. Now consider rapidly raising the well depth such that $J \approx 0$. The system retains memory of its initial state, and this state time evolves with the zero-hopping Hamiltonian, $H_{final}$. We phenomenologically include the dissipation through the cavity walls, parametrized by $\kappa$. Thus, the system time evolves with the following effective Hamiltonian after the quench: 
\bea
H_{final} &=& \sum_i (\epsilon_{a} n_{ia} + \epsilon_{b} n_{ib}+\omega \psi_{i}^{\dagger} \psi_{i} +  \\ \nonumber
&&+g (b_{i}^{\dagger} a_{i} \psi_{i} + b_i a_i^{\dagger} \psi_{i}^{\dagger})- \mu_{1} n_{i1} - \mu_{2} n_{i2} - i \frac{\kappa}{2} \psi_i^{\dagger} \psi_i)
\eea
where $\kappa \ll g$ because the system is assumed to be strongly coupled.  

Figure \ref{timeEvolPh}a shows the average intracavity photon number density as a function of time after the quench. Both in the absence of dissipation (red) and in the presence of dissipation (blue) we see oscillatory photon number.  We accounted for dissipation by time evolving with wave function Monte Carlo \cite{MolmerPaper, QMJumpsReview}. The probability of detecting a photon outside the cavity is directly proportional to the mean intracavity photon number; thus these intracavity oscillations are expected to result in oscillatory photon counting outside the cavity.

To understand the oscillatory behavior, we write the initial SF state of the system as a linear combination of the eigenstates of $H_{final}$:
\bea
\left| S_{initial} \right> &=& c_0 e^{-i E_{0} t} \left| 0,0,0 \right>+\\ \nonumber
&&+c_1 e^{-i E_{1} t} \left| 0,0,1 \right>+\\ \nonumber
&&+\sum_{k=1}^{k=k_{max}}\Big[d_k e^{-i E_{k11} t} \left| k,1,1 \right>+\\ \nonumber
&&+c_{k1} e^{-i E_{-}^{(k)} t} \left| - , n_{ph}=k \right>+\\ \nonumber  
&&+c_{k2} e^{-i E_{+}^{(k)} t} \left| + , n_{ph}=k \right>\Big]
\eea
where $| n_{ph}=k,  - \rangle$  and $| n_{ph}=k, + \rangle$ are the lower and upper branch polariton states with k excitations. $E_{-}^{(k)}$ and $E_{+}^{(k)}$ are their respective energies, and $k_{max}$ is the maximum photon number. Furthermore, the Fock states ($\left| 0,0,1 \right>$ and $\left| k,1,1 \right>$) denote $\left| n_{ph},n_b,n_a \right>$. From here, $\langle n_{ph} \rangle(t)$ is simply:
\be
\langle n_{ph} \rangle (t) = A_0 - \sum_{k=1}^{k=k_{max}} A_k \cos(\Omega_k t)
\ee
with $\Omega_k=E_{+}^{(k)}-E_{-}^{(k)}$. Therefore each frequency is the energy difference between the upper and lower polariton states with $k$ excitations. Further, we can see that oscillation occurs only when the initial state has nonzero weight in \textit{both} of these polariton states; more precisely, both $c_{k1}$ and $c_{k2}$ must be nonzero in order for the number to oscillate at frequency $\Omega_k$. For the particular parameters of Figure \ref{timeEvolPh}, the initial (mean field) state of the system has weight in the following states:  $| n_{ph}=1,  - \rangle$, $| n_{ph}=1, + \rangle$, $| n_{ph}=2,  - \rangle$, and $| n_{ph}=2, + \rangle$. Therefore, two frequencies are present in the quench dynamics (Figure \ref{timeEvolPh}b). The dominant frequency corresponds to the $k=1$ polariton, and the subtler frequency corresponds to the $k=2$ polariton. 

Next we explore how the dynamics change as the initial order parameters change. Consider a system with low initial $J$ such that the initial system is Mott insulating. As noted in the previous section, the initial state here is simply a polariton on each site, $| S_0 \rangle = \prod_{i=1}^N | n_{ph}, - \rangle_i$. If there is zero dissipation, this state is just an eigenstate of $H_{final}$, so all amplitudes are zero. In the presence of dissipation, there is a small offset from zero.   

Now consider starting in the superfluid regime. At smaller initial $J$, the initial state has more weight concentrated in one particular polariton state over the others. That is, from the set of $\{c_{k1}, c_{k2}\}$ one particular mode
dominates over the others. For larger initial $J$, the initial state has its weight diffused over multiple polariton states, so all the modes from the set $\{c_{k1}, c_{k2}\}$ contribute. The validity of these statements is reflected in Figure ~\ref{eqplots}c, where $| \langle \psi_i b_i^{\dagger} a_i \rangle |$ decreases with $\frac{zJ}{g}$ as the nonlocal kinetic term increasingly dominates over the local atom-photon interaction. Thus, if we start in the superfluid regime, the amplitudes, $A_k \propto c_{k1}c_{k2}$, scale with J.  

Figure \ref{timeEvolPh}c exactly exhibits this correspondence between the order parameter and the amplitudes of oscillation. The amplitudes, $A_1$ and $A_2$, are extracted by fitting the Monte Carlo curves to the following:
\be \label{nphvst}
\langle n_{ph} \rangle (t) = A_0 - ( A_1 \cos(\Omega_1 t)+ A_2 \cos(\Omega_2 t))e^{-c t}
\ee

We emphasize that Figure \ref{timeEvolPh}c relates to the dynamics of the intracavity photon number $\textit{density}$. The total \textit{number} within the cavity scales with the number of lattice sites. As a result, the amplitude trends in Figure \ref{timeEvolPh}c would be far more pronounced in actual experiment. 

Thus, we come to the central conclusion of this section; the correspondence between the initial order parameters and the photon number quench dynamics allows elucidation of the initial quantum phase from photon statistics alone.

\textit{Experimental Method to Determine Superfluid Order Parameters.--} Our theoretical analysis indicates that photon number measurement allows detection of superfluidity in a two band Bose Hubbard system coupled to light. Furthermore, the light-matter interaction bridges the photon number dynamics and the superfluid order parameters. As a result, measuring the photon number amplitudes of oscillation after a quench provides an easy experimental measure of the order parameter. Here we note that we expect realization of our calculations to be possible within the near future. In particular, the group of Thompson and collaborators \cite{Thompson} succeeded in placing an optical lattice in a cavity, albeit a thermal system.

Future directions for this work can treat dissipative effects more carefully. For our calculations we assumed $\frac{g}{\kappa}=20$ and negligible decay of the upper state, $\gamma$. Experiments may benefit from knowing how $\gamma$ and $\kappa$ affect the above trends.  

%We gratefully acknowledge useful discussions with James Thompson (JILA). 
\textit{Acknowledgements.--}This work was supported by grant NSF DMR 0907275. SR acknowledges support from the NSF Graduate Research Fellowship Program.

\end{document}